\documentclass[numreferences]{kluwer}

\usepackage{bbm,mathrsfs}

\newcommand{\matice}[1]{\left( \begin{array}{cc} #1 \end{array} \right)}
\newcommand{\eq}[1]{\begin{equation} #1 \end{equation}}
\newcommand{\eqarray}[2]{ \begin{eqnarray}  #1  \label{#2} \end{eqnarray} }
\newcommand{\<}{\langle}
\renewcommand{\>}{\rangle}
 
\newcommand{\R}{{\mathbbm{R}}}
\newcommand{\C}{{\mathbbm{C}}}

\newcommand{\T}{{\mathcal{T}}}
\newcommand{\dd}{{{\rm d}}}
\newcommand{\ii}{{\rm i}}
\renewcommand{\P}{{\mathcal{P}}}
\newcommand{\PT}{{\mathcal{PT}}}
\renewcommand{\H}{{\mathcal{H}}}
\newcommand{\Dom}{{\rm{Dom}\,}}
\newcommand{\Ker}{{\rm{Ker}}}
\newcommand{\Ran}{{\rm{Ran}}}

\newcommand{\wlim}{\mathop{\mathrm{w}\mbox{--}\lim}}

\renewcommand{\Re}{{\rm Re}\,}
\renewcommand{\Im}{{\rm Im}\,}
\newcommand{\BH}{\mathscr{B}(\mathcal{H})}

\begin{document}
\begin{article}
\begin{opening}
\title{Surprising spectra of $\PT$-symmetric point interactions }

\author{Petr \surname{Siegl}\email{siegl@ujf.cas.cz}}  
\institute{Laboratoire Astroparticules et Cosmologie, Universit\'e Paris 7, Paris, France,\\
Department of Theoretical Physics, Nuclear Physics Institute, Academy of Sciences, \v Re\v z, Czech Republic\\
						Faculty of Nuclear Sciences and Physical Engineering, Czech Technical University, Prague, Czech Republic.}

\begin{abstract} 
Spectra of the second derivative operators corresponding to the special $\PT$-symmetric point interactions are studied. The results are partly the completion of those obtained in \cite{albeverio-2002-59}. The particular $\PT$-symmetric point interactions causing unusual spectral effects are investigated for the systems defined on a finite interval as well. The spectrum of this type of interactions is very far from the self-adjoint case despite of $\PT$-symmetry, $\P$-pseudo-Hermiticity and $\T$-self-adjointness.
\end{abstract} 


\classification{Mathematics Subject Classification (2000)}{47A55, 47B99, 81Q05}

\end{opening}
\newdisplay{define}{Definition}
\newtheorem{prop}{Proposition}
\newtheorem{thm}{Theorem}
\newtheorem{rem}{Remark}
\newtheorem{cor}{Corollary}
\newtheorem{lem}{Lemma}
\newdisplay{example}{Example}

\section{Introduction}

$\PT$-symmetric operators, a special case of operators with antilinear symmetry, have been intensively studied in both physical and mathematical context as a result of the observation that the spectrum of such operators may be real and discrete \cite{bender-1998-80}. Although it is known that some $\PT$-symmetric operators are special case of quasi-Hermitian ones \cite{Dieudonne1961}, or equivalently, they can be mapped by similarity transformation to the self-adjoint ones, see e.g. \cite{Albeverio2005-38,krejcirik-2006-39,Siegl2008-41} for examples, the spectrum of $\PT$-symmetric operators may be also complex, e.g. complex conjugated eigenvalues may appear already for matrices with antilinear symmetry. The complex conjugated pairs of eigenvalues instead of the real ones are actually the simplest possible deviation of the spectrum from the self-adjoint case. In fact, the class of operators with antilinear symmetry is much larger. The residual spectrum of operators (even bounded) with antilinear symmetry  may be non-empty and the point spectrum of such operator may be uncountable \cite{Siegl2009-Pr}, i.e. operators may be non-spectral \cite{DSIII}. 

Our aim is to present more accurate results for particular $\PT$-symmetric point interactions on a line, described in general in \cite{albeverio-2002-59}. These differential operators exhibits very interesting spectral properties, the spectrum can be entire complex plane with uncountable point spectrum instead the two complex conjugated eigenvalues appearing usually \cite{albeverio-2002-59}. We show that the spectrum of analogous models defined on the finite interval $(-l,l)$ is either empty or entire complex plane depending on boundary conditions imposed at $\pm l$. In the physical framework, the fact that these point interactions can completely and dramatically change the spectrum is surprising. Nonetheless, considering operators being not even similar to the normal ones brings expected unusual spectral effects. From this point of view, the simplicity of the presented examples may be credited. The examples defined on the finite interval emphasize the necessity of the non-empty residual set assumption in \cite[III, Corollary 6.34]{Kato}. The claim of this corollary is essentially that the extension of finite order has the compact resolvent if and only if some other extension of the same operator has the compact resolvent.

We find explicitly the boundary conditions for the adjoint operators in the first section and we also put slightly more precisely the claim of \cite{albeverio-2002-59} concerning the $\PT$-self-adjointness of the operators. The proofs of the closedness of operators are based on the relation to the adjoint operator as well. In the next section, we investigate the particular $\PT$-symmetric point interaction for the model defined on a line and we present the results on the spectrum and relation to the collapse of quasi-Hermiticity. Models defined on the finite interval $(-l,l)$ are studied in the last section. The dependence of the spectrum on the boundary conditions at $\pm l$ is described in details.

$\PT$-symmetry is defined in $L^2(\R)$ spaces in the following way, the parity $\P$ acts as $(\P\psi)(x)=\psi(-x)$ and the time reversal symmetry $\T$ is the complex conjugation $(\T\psi)(x)=\overline{\psi}(x)$. We say that an operator $A$ is $\PT$-symmetric if $\PT A \psi = A \PT \psi$ for all $\psi \in \Dom (A)$.

In order to avoid any confusion, we recall the definition of the division of the spectrum for the closed operator $A$ in a Hilbert space $\H$ which we use and can be found e.g. in \cite{BEH}. A complex number belonging to the spectrum of $A$ is \\
$i)$ in the point spectrum $\lambda\in\sigma_p(A)$ if $\Ker(A-\lambda)\neq \{0 \}$,\\
$ii)$ in the continuous spectrum $\lambda\in\sigma_c(A)$  if $\Ker(A-\lambda)= \{0 \}$ and $\overline{\Ran(A-\lambda I)}=\H$,\\
$iii)$ in the residual spectrum $\lambda\in\sigma_r(A)$  if $\Ker(A-\lambda)= \{0 \}$ and $\overline{\Ran(A-\lambda I)}\neq\H$.

\section{$\PT$-symmetric point interactions - adjoint operator} 

We consider family of $\PT$-symmetric point interaction at the origin determined in \cite{albeverio-2002-59} by the two types of boundary conditions - connected and separated. Differential operator $L$ corresponding to the point interaction
\eq{L=-\frac{\dd^2}{\dd x^2} }
is defined on the domain $\Dom(L)$ consisting of the functions $\psi$ from $W^{2,2}(\R\setminus\{0\})$ satisfying boundary conditions described by parameters $b,c,\psi,\theta,h_0,h_1$ in the following way
\begin{itemize}
\item[i)] connected case

\eq{\matice{\psi(0+) \\ \psi'(0+)}=B \matice{\psi(0-) \\ \psi'(0-)}, \label{PTPBound}}
with the matrix $B$ equal to
\eq{B=e^{\ii\theta}\matice{\sqrt{1+bc}e^{\ii\phi} & b \\ c & \sqrt{1+bc}e^{-\ii\phi}}, \label{Bmatrix} }
with the real parameters $b\geq 0$, $c \geq -1/b$, $\theta,\phi \in (-\pi,\pi]$.

\item[ii)] separated case
\eqarray{	h_0 \psi'(0+) &=& h_1 e^{\ii\theta} \psi (0+), \nonumber \\
   				h_0 \psi'(0-) &=& -h_1 e^{-\ii\theta} \psi (0-),
   			}{SepPTCon}
with the real phase parameter $\theta \in [0,2\pi)$ and with the parameter {\bfseries h} $=(h_0,h_1)$ taken from the (real) projective space {\bfseries P}$^1$.
\end{itemize}

The operator $L$ is an extension of a symmetric densely defined operator $L_0=-\dd^2/ \dd x^2$
with the domain $\Dom(L_0)=C_0^{\infty}(\R\setminus\{0\})$. $L$ can be also viewed as a restriction of $L_{max}=L_0^*=-\dd^2/ \dd x^2$ with the domain $\Dom(L_{max})=W^{2,2}(\R\setminus\{0\})$.

At first, we determine the adjoint operator $L^*$ explicitly.

\begin{prop}

Let $L$ be the second derivative operator corresponding to the $\PT$-symmetric point interaction (\ref{PTPBound}-\ref{SepPTCon}). The adjoint operator $L^*$ is the second derivative operator defined on the domain $\Dom(L^*)$ including functions $\varphi$ from  $W^{2,2}(\R\setminus\{0\})$ which satisfy following boundary conditions

\begin{itemize}
\item[i)] connected case

\eq{\matice{\varphi(0-) \\ \varphi'(0-)}=\tilde{B} \matice{\varphi(0+) \\ \varphi'(0+)}, \label{PTPBoundAdj}}
with the matrix $\tilde{B}$ equal to
\eq{\tilde{B}=e^{-\ii\theta}\matice{\sqrt{1+bc}e^{\ii\phi} & -b \\ -c & \sqrt{1+bc}e^{-\ii\phi}}, \label{BmatrixAdj} }
with the real parameters $b\geq 0$, $c \geq -1/b$, $\theta,\phi \in (-\pi,\pi]$.

\item[ii)] separated case
\eqarray{	h_0 \varphi'(0+) &=&  h_1 e^{-\ii\theta} \varphi (0+), \nonumber \\
   				h_0 \varphi'(0-) &=& -h_1 e^{ \ii\theta} \varphi (0-),
   			}{SepPTConAdj}
with the real phase parameter $\theta \in [0,2\pi)$ and with the parameter {\bfseries h} $=(h_0,h_1)$ taken from the (real) projective space {\bfseries P}$^1$.
\end{itemize}
\label{adjoint}
\end{prop}

\begin{pf}

Since the operator $L$ is the extension of $L_0$, the adjoint $L^*$ is the restriction of $L_0^*=L_{max}$ by the very well known relation between operator and its adjoint $A\subset B \Rightarrow B^* \subset A^*$. Hence $L^*$ acts as the second derivative operator in $W^{2,2}(\R\setminus\{0\})$ and it remains to determine the boundary conditions only. We consider a function $\varphi\in\Dom(L^*)$, the definition of the adjoint operator yields the equality
\eq{\int_{\R}\overline{\varphi}(x)\psi''(x)\dd x=\int_{\R}{\varphi}''(x)\psi(x)\dd x, \label{AdjDefEq}}
for all $\psi \in \Dom(L)$. By using integration by parts for the left-hand side of (\ref{AdjDefEq}) and inserting boundary conditions (\ref{PTPBound}-\ref{SepPTCon}) for $\psi \in \Dom(L)$ we obtain
\eqarray{\overline{\psi}'(0-)\left[ \varphi(0-)- \varphi(0+)e^{\ii(\phi-\theta)}\sqrt{1+bc} + \varphi'(0+)e^{-\ii\theta}b \right] + &&\\ \nonumber
	+	\overline{\psi}(0-) \left[ -\varphi'(0-)-\varphi(0+)e^{-\ii\theta}c+\varphi'(0+)e^{-\ii(\phi+\theta)}\sqrt{1+bc} \ \right]&=&0 }{L*B1}
for the connected case and
\eqarray{\overline{\psi}(0+)h_0 \left[h_0\varphi'(0+)-h_1 e^{-\ii\theta}\varphi(0+)  \right] - &&\\ \nonumber
-\overline{\psi}'(0-)h_0\left[h_0\varphi'(0-)+h_1 e^{\ii\theta}\varphi(0+)  \right]  &=&0 }{L*B1s}
for the separated case. Since the boundary conditions for $\psi\in\Dom(L)$ have been already applied, the values of $\psi(0-),\psi'(0-)$ for the connected case and $\psi(0\pm)$ for separated case are arbitrary and hence $\varphi$ must satisfy boundary conditions (\ref{PTPBoundAdj}-\ref{SepPTConAdj}).
\begin{flushright}
$\square$
\end{flushright}
\end{pf}

We would like to remark that the claim of \cite{albeverio-2002-59} that all operators $L$ satisfy the property $L^*=\P L\P$ is not entirely accurate for the connected case. The domain of $\P L \P$ is the $\P^{-1}=\P$ image of the $\Dom(L)$, i.e.
\eq{\psi \in \Dom(\P L) \Leftrightarrow \psi \in \P \Dom(L) \Leftrightarrow \P\psi\in \Dom(L)\Leftrightarrow \nonumber} 
\eq{\Leftrightarrow\matice{\psi(0-) \\ \psi'(0-)}=\hat{B} \matice{\psi(0+) \\ \psi'(0+)}, \label{PDomL}}
where $\hat{B}=e^{2\ii\theta}\tilde{B}$. Thus, the relation $L^*=\P L\P$, technically the equality of $\hat{B}$ and $\tilde{B}$, is valid only for $\theta=0$. Nevertheless, none of the other claims of \cite{albeverio-2002-59} is affected by this fact because of the unitary equivalence of the operators corresponding to the different choices of $\theta$. We will consider only $\theta=0$ further.

We summarize symmetry properties of $L$. The proof of the following proposition is straightforward application of boundary conditions for $L,L^*$ and actions of operators $\P$ and $\T$.
\begin{prop}
Let $L$ be the second derivative operator corresponding to the connected $\PT$-symmetric point interaction at the origin (\ref{PTPBound},\ref{Bmatrix}) with the choice $\theta=0$ in the boundary conditions. Then \\ 
i) $L^*=\P L \P$, \\
ii) $\forall \psi \in \Dom(L), \ \ \PT L \psi = L \PT \psi,$ \\
iii) $L^*=\T L \T.$
\label{sym}
\end{prop}
The first symmetry is referred to as the $\P$-pseudo-Hermiticity or $\PT$-self-adjointness, the second one is $\PT$-symmetry in its original sense and the third one is the $\T$-self-adjointness, the special case of $J$-self-adjointness, where $J$ is an antilinear isometric involution, i.e. $J^2 = I$ and $\<Jx, Jy\> = \<y, x\>$ for all $x,y \in \H$. The importance of $\T$-self-adjointness for $\PT$-symmetric models was stressed in \cite{borisov-2007}, one of the reasons is that the residual spectrum of $J$-self-adjoint operators is empty \cite[Lem. III.5.4]{EE}.

We would like to stress that the property $i)$ of the proposition \ref{sym} guarantees that the operator $L$ is closed. To this end take into the consideration closedness of every adjoint operator, the relation $i)$, and $\P\in \BH$. We demonstrate the closedness of the operators for particular models the operator $L$ in following sections.

\section{Model on a line}

Spectrum of the $\PT$-symmetric point interactions has been investigated in \cite[Thm.2, Prop.1]{albeverio-2002-59}. It basically consists of the branch of continuous spectrum $[0,\infty)$ and up to two real or complex conjugated eigenvalues. We demonstrate that for the special case of the interaction being described in the following the characterization of the spectrum is different. 

Let us study the connected case with $\theta=b=c=0$, i.e. boundary conditions for $L_{\phi}$ read
\eq{\psi(0+)=e^{\ii \phi}\psi(0-), \ \ \psi'(0+)=e^{-\ii \phi}\psi'(0-), \label{BCSpCase}}
where $\phi \in (-\pi,\pi].$ The case $\phi=\pi$ corresponds to the self-adjoint operator.
The adjoint operator $L^*_{\phi}$ is given by Proposition \ref{adjoint}, $L^*_{\phi}=L_{-\phi}$ in fact. This relation also proves that $L_{\phi}$ is a closed operator considering the closedness of every adjoint operator. 

Spectral properties of $L_{\phi}$ for $\phi\neq \pm\frac{\pi}{2}$ are very simple, the spectrum is continuous without any eigenvalues, 
\eq{\sigma(L_{\phi})=\sigma_c(L_{\phi})=[0,\infty),  \ \ \ \phi\neq \pm \frac{\pi}{2}. \label{SpLUsual} }
It is possible to find an invertible positive bounded operator $\Theta$ with bounded inverse satisfying
\eq{L_{\phi}^* \Theta_{\phi}=\Theta_{\phi} L_{\phi}, \ \ \ \phi\neq \pm \frac{\pi}{2}, \label{QH}  }
in other words, to show that $L_{\phi}$ is quasi-Hermitian \cite{Dieudonne1961} or equivalently that $L_{\phi}$ is similar to the self-adjoint operator. The explicit formula for the operator $\Theta$ and its square root was obtained by different approaches in \cite{Albeverio2005-38,Siegl2008-41,Albeverio2009-42},
\eq{\Theta_{\phi}=I - \ii \sin\phi P_{\rm sign}\P, \label{Theta}}
where the operator $P_{\rm sign}$ acts as the multiplication by the function ${\rm sign}\, x$. The spectrum of $\Theta_{\phi}$ consists of the two eigenvalues $1\pm\sin\phi$ of infinite multiplicities,
\eq{\sigma(\Theta_{\phi})=\sigma_p(\Theta_{\phi})=\{1\pm\sin\phi\}.} 

We denote $L_{\pm}, \Theta_{\pm}$ operators corresponding to $\phi=\pm \frac{\pi}{2}$.
The relation (\ref{QH}) is still valid for $\phi=\pm \frac{\pi}{2}$, however, operators $\Theta_{\pm}$ are no longer invertible. Moreover, we can see that formula for the resolvent \cite[eq.(17)]{albeverio-2002-59} collapses because the \cite[eq.(18)]{albeverio-2002-59} appearing in the denominator of $\rho_{\pm}$ is identically zero. These facts are reflected in unusual spectral properties of $L_{\pm}$ being far from those of self-adjoint operators. 
\begin{prop}
Spectra of the operators $L_{\pm}$ include all complex numbers. The residual part of the spectra is empty, $[0,\infty)$ is the continuous part and every $\lambda \in \C\setminus [0,\infty)$ belongs to the point spectrum.
\eq{\sigma_p(L_{\pm})=\C\setminus [0,\infty), \ \ \sigma_c(L_{\pm})=[0,\infty), \ \ \sigma_r(L_{\pm})=\emptyset.}
\end{prop}

\begin{pf}
The residual spectrum is empty, as it was mentioned before, because of the $\T$-self-adjointness of all $L_{\phi}$, \cite[Lem. III.5.4]{EE} and the equality  $[\Ran(A)]^\bot = \Ker(A^*)$ valid for every densely defined operator $A$, see \cite[Cor. 2.1.]{borisov-2007} for the detailed discussion.

We construct an eigenfunction for every $\lambda \in \C\setminus[0,\infty)$. We define functions $\psi_{k\pm},\varphi_{k\pm},\zeta_{k\pm}$,
\eqarray{
\psi_{k\pm}(x)&=&
\left \{
\begin{array}{rl}
				e^{kx}, 						& x < 0, \\
				\pm\ii e^{-kx}, 				& x > 0,
  \end{array} \right. \ \ 
\varphi_{k\pm}(x)=
\left \{
\begin{array}{rl}
				e^{-kx}, 						& x < 0, \\
				\pm\ii e^{kx}, 				& x > 0,
  \end{array} \right. \\ 
  \zeta_{k\pm}(x)&=&
\left \{
\begin{array}{rl}
				e^{-\ii kx}, 						& x < 0, \\
				\pm\ii e^{\ii kx}, 				& x > 0.
  \end{array} \right.  
				}{EigenFunct}
These functions satisfy the equations 
\eq{L_{\pm}\psi_{k\pm}=-k^2\psi_{k\pm}, \ \ L_{\pm}\varphi_{k\pm}=-k^2\varphi_{k\pm}, \ \ L_{\pm}\zeta_{k\pm}=k^2\zeta_{k\pm},}
and the boundary conditions for the domains of $L_{\pm}$ as well. $\psi_{k\pm}$ are in $L^2(\R)$ for $\Re k>0$, $\varphi_{k\pm}$ for $\Re k<0$ and 
$\zeta_{k\pm}$ for $\Re k=0$ and $\Im k >0$.

The interval $[0,\infty)$ is not in the point spectrum because corresponding solutions of the equation $L_{\pm}f=\lambda f$ are not in $L^2(\R)$. Since the spectrum is a closed set, the interval $[0,\infty)$ is in the spectrum. To be more precise, due to the disjoint decomposition of the spectrum, the interval belongs to the continuous part. 

Alternatively, we can prove that the interval $[0,\infty)$ is in the spectrum by using the Weyl criterion. Inspired by \cite[Proof of Lem.5.3.]{krejcirik-2005-41}, we consider a sequence of functions from $\Dom(L_{\pm})$
\eq{ \hat{\psi}_n(x)= e^{\ii kx} \varphi_n(x), }
where $\varphi_n(x):=\varphi(x/n -n) $ with $\varphi\in C_0^{\infty}\left( (-1,1)\right)$, $k\in\R$ . We introduce normalized functions $\psi_n=\hat{\psi}_n/\|\hat{\psi}_n\|$. Every $\psi_n$ is in the $\Dom(L_{\pm})$ because the support of $\psi_n$ does not contain zero. We check that  \eq{\wlim_{n\rightarrow \infty} \psi_n=0, \label{w0}} 
i.e. $\<\chi,\psi_n\>\rightarrow 0$ for all $\chi \in L^2(\R)$. Since $\|\psi_n\|=1$, it is sufficient to show (\ref{w0}) for all $\chi \in C_0^{\infty}(\R)$ only, the remaining follows by a standard density argument. However, it is obvious that the weak limit is zero for such a $\chi$ because the supports of $\chi$ and $\psi_n$ are disjoint for $n$ large enough. Next,
\eq{\|(L_{\pm}-k^2)\psi_n\|\rightarrow 0,}
for all $k\in\R$ by virtue of
\eq{\|(L_{\pm}-k^2)\psi_n\|\leq 2|k|\frac{\|\varphi_n'\|}{\|\varphi_n\|} + \frac{\|\varphi_n''\|}{\|\varphi_n\|} }
and the definition of the sequence $\{\varphi_n\}$. Thus we showed that $k^2$ in the essential spectrum for every $k\in\R$ by using the Weyl criterion.
\begin{flushright}
$\square$
\end{flushright}
\end{pf}

\section{Models on a finite interval}

We consider finite interval $(-l,l)$ and second derivative operator $L_{\phi}$ corresponding to the $\PT$-symmetric interaction at origin of the type (\ref{BCSpCase}). The domain of $L_{\phi}$ consists of functions $\psi$ belonging to the Sobolev space $W^{2,2}((-l,0)\cup(0,l))$ and satisfying boundary conditions (\ref{BCSpCase}) at origin and some other boundary conditions at $\pm l$ being specified later. Our aim is to study the spectrum of such differential operators, particularly for the $\phi=\pm \pi/2$ case. We show that the choice of the boundary conditions at $\pm l$ plays an essential role. If we slightly modify the proof of the proposition \ref{adjoint}, we can easily obtain formula for the adjoint operators and prove that considered operators are closed. 

We distinguish to classes of boundary conditions being imposed at $\pm l$, symmetric and $\PT$-symmetric ones. Symmetric boundary conditions are determined by a unitary matrix $U$ entering well known relation 
\eq{(U-I)\Psi(l) + {\ii}\,L_0(U+I)\Psi'(l)=0, \label{SymBC}}
where $L_0\in \R$ and 
\eq{ \Psi(l)=\matice{\psi(l) \\ \psi(-l)}, \ \ \ \ \Psi'(l)=\matice{\psi'(l) \\ -\psi'(-l)} \label{PsiPsi0'}.}
$\PT$-symmetric boundary conditions, previously already discussed, are defined by relations (\ref{PTPBound}-\ref{SepPTCon}). 

We summarize spectral properties of $L_\phi$ in following propositions. As we may expect, the case $\phi=\pm \pi/2$ exhibits unusual features.
\begin{prop}
Let $L_{\phi}$ be the second derivative operator in $L^2((-l,l))$ corresponding to the $\PT$-symmetric point interaction (\ref{BCSpCase}) at origin with symmetric boundary conditions (\ref{SymBC}-\ref{PsiPsi0'}) at $\pm l$. 

Let $\phi \neq \pm \pi/2$. Then the spectrum of $L_{\phi}$ is discrete and its eigenvalues $\lambda=k^2$ are solutions of the equation 
\eqarray{ 
\cos\phi \Big( P_1(U) -2 \ii k L_0 P_2(U)\cos 2kl +k^2 L_0^2 P_3(U) \sin 2kl  \Big) +\nonumber \\
+ 2 \ii k L_0 \Big(u_{12}+u_{21}+\ii (u_{11}-u_{22}) \sin\phi\Big) =0,}{SpSym}
where $u_{ij}$ are elements of the unitary matrix $U$ (\ref{SymBC}) and
\eqarray{
P_1(U)&=&1-u_{11}-u_{12} u_{21}-u_{22}+u_{11} u_{22},\nonumber \\
P_2(U)&=&1+u_{12} u_{21}-u_{11} u_{22},\nonumber \\
P_3(U)&=&1+u_{11}-u_{12} u_{21}+u_{22}+u_{11} u_{22}.
}{PU}

Let $\phi = \pm \pi/2$. Then the point spectrum of $L_{\pm}$ is either empty or entire $\C$. The latter case occurs if and only if 
\eq{u_{12}+u_{21}\pm \ii (u_{11}-u_{22}) = 0. \label{EmptySym}}

\end{prop}

If we take into consideration usual Dirichlet ($U=-I$), Neumann ($U=I$) or Robin ($U=\alpha I, \alpha\in \R$) boundary conditions at $\pm l$, then the condition (\ref{EmptySym}) is fulfilled, thus the spectrum of $L_{\pm}$ is entire complex plane. 

Next, we apply both connected and separated $\PT$-symmetric boundary conditions at $\pm l$. It may be expected for connected case that the second point interaction (parameters are denoted by the subscript 2) of the type $b_2=0,c_2=0,\phi_2 = \pm \pi/2$ produces analogous interesting spectral effects. 

\begin{prop}
Let $L_{\phi}$ be the second derivative operator in $L^2((-l,l))$ corresponding to the $\PT$-symmetric point interaction (\ref{BCSpCase}) at origin with connected $\PT$-symmetric boundary conditions (\ref{PTPBound}) at $\pm l$. 

Let $\phi \neq \pm \pi/2,\phi_2\neq \pm \pi/2$ or $\phi \neq \pm \pi/2, \phi_2= \pm \pi/2 $ and $b_2\neq0$ or $c_2\neq0$. Then the spectrum of $L_{\phi}$ is discrete and its eigenvalues $\lambda=k^2$ are solutions of the equation 
\eqarray{\cos\phi\left( \left(b_2 k^2-c_2\right) \sin 2kl+2 k \sqrt{1+b_2 c_2} \cos \phi_2 \cos 2kl \right)+\nonumber\\
+2 k \left(\sqrt{1+b_2 c_2} \sin\phi \sin \phi_2-1 \right)=0.
}{PSymPTCon}

Let $\phi = \pm \pi/2$. Then the point spectrum of $L_{\pm}$ is either empty or entire $\C$. The latter case occurs if and only if 
\eq{ \sqrt{1+b_2 c_2} \sin\phi_2-1 =0.}

Let $b_2=0,c_2=0,\phi_2= \pm \pi/2$ then the point spectrum of $L_{\pm}$ is either empty or entire $\C$. The latter case occurs if and only if $\phi=\pm\pi/2$.

\end{prop}

\begin{prop}
Let $L_{\phi}$ be second derivative operator in $L^2((-l,l))$ corresponding to the $\PT$-symmetric point interaction (\ref{BCSpCase}) at origin with separated $\PT$-symmetric boundary conditions (\ref{SepPTCon}) at $\pm l$.  

Let $\phi \neq \pm \pi/2$ and $\theta \neq 0, \pi$. Then the spectrum of $L_{\phi}$ is discrete and its eigenvalues $\lambda=k^2$ are solutions of the  equation 
\eqarray{\cos \phi \left(2 h_0 h_1 k \cos 2kl \cos \theta+\left(h_0^2 k^2-h_1^2\right) \sin 2kl\right)- \nonumber \\
-2 h_0 h_1 k \sin \theta \sin \phi=0.}{PTSepImpl}

Let $\phi = \pm \pi/2$. Then the point spectrum of $L_{\pm}$ is either empty or entire $\C$. The latter case occurs if and only if $\theta=0,\pi$, i.e. separated conditions are symmetric Robin conditions.
\label{last}
\end{prop}

\begin{rem}
The case of empty point spectrum actually means that the whole spectrum is empty because the resolvent is compact in this case.
\end{rem}

\begin{pf}
We solve the eigenvalue problem $L_{\phi}\psi=\lambda\psi$ together with both boundary conditions. We search for the non-zero eigenfunction and this is reproduced in terms of the secular equations (\ref{SpSym},\ref{PSymPTCon},\ref{PTSepImpl}). If we insert there $\phi=\pm \pi/2$ or other assumptions on the rest of parameters, we obtain the assertions concerning the empty and entire $\C$ point spectrum. 

In order to prove the claim of the non-empty discrete spectrum and of the remark above we show that the resolvent is compact in these cases. We calculate the resolvent explicitly for the operator $L_+$ in the proposition \ref{last}. The remaining resolvents can be obtain by analogous procedure.
At first, using standard Green function approach we calculate the resolvent corresponding to the $L_1=-\dd^2 / \dd x^2$ on $(-l,l)$ with separated $\PT$-symmetric conditions (\ref{SepPTCon}) at $\pm l$.
\eq{\left(R_{L_1}(\lambda)g\right)(x)=\int_{-l}^l G(x,y) g(y) \dd y,  \label{R1} }
where $g \in L_2(\R)$, $\lambda=k^2$, and 
\eq{G(x,y)= \frac{1}{W(k)} 
\left\{\begin{array}{ll}
 	u_-(x)u_+(y), & x\leq y\\
				u_-(y)u_+(x), & x\geq y,															
\end{array}
  \right.  }
\eqarray{W(k)&=&\frac{k}{h_0^2} (-2 h_0 h_1 k \cos 2kl \cos\theta+(h_1-h_0 k) (h_1+h_0 k) \sin 2kl) \nonumber\\
				u_-(x)&=&\cos kx \left(-k \cos kl+e^{-\ii \theta} \frac{h_1}{h_0} \sin kl\right)+\left(e^{-\ii \theta } \frac{h_1}{h_0}  \cos kl+k \sin kl\right) \sin kx, \nonumber \\
				u_+(x)&=&	\cos kx \left(k \cos kl-e^{\ii \theta } \frac{h_1}{h_0} \sin kl \right)+\left(e^{\ii \theta } \frac{h_1}{h_0} \cos kl+k \sin kl\right) \sin k x.		}{wronskian}
We may easily check that functions $u_{\pm}$ satisfy appropriate boundary condition (\ref{SepPTCon}) at $\pm l$. We proceed by determining the basis of $\Ker(L_1^*-\lambda)$ which we denote $e_{\pm}$, 
\eqarray{e_-(x)&=& \left\{ 
\begin{array}{ll}
\left(-k \cos kl+e^{\ii \theta } \frac{h_1}{h_0}  \sin kl\right)\cos k x + & \\ 
+\left(e^{\ii \theta } \frac{h_1}{h_0}  \cos kl+k \sin kl \right) \sin k x, & x\leq 0, \\
0, & x>0,
\end{array} \right.
\nonumber\\
				e_+(x)&=&\left\{ 
\begin{array}{ll}
0, & x<0,\\
\left(k \cos kl-e^{-\ii \theta } \frac{h_1}{h_0}  \sin kl\right)\cos k x +& \\
+\left(e^{-\ii \theta } \frac{h_1}{h_0}  \cos kl+k \sin kl \right) \sin kx, & x\geq 0.				
\end{array} \right.				
				}{epm}
Then the resolvent of $L_{\phi}$ can be written in the form
\eq{ \left(R_{L_{\phi}}(\lambda)g \right)(x)=\left(R_{L_1}(\lambda)g\right)(x) + C_-(k)e_-(x)+C_+(k)e_+(x),} 
where constants $C_{\pm}$ are to be determine. $R_{L_{\phi}}(\lambda)g \in \Dom L_{\phi},$ thus it must satisfy boundary conditions (\ref{BCSpCase}). This leads to the system of linear equations for $C_{\pm}$
\eqarray{M \matice{C_- \\ C_+} = \matice{ \left(e^{\ii \phi }-1\right) F_1(0) \\
																					\left(e^{-\ii \phi }-1\right) F_1'(0)}, }{EqCpm}
where
\eqarray{M&=&
\matice{e^{\ii \phi } \left(k \cos kl -e^{\ii \theta} \frac{h_1}{h_0}  \sin kl\right) &  k\cos kl -e^{-\ii \theta } \frac{h_1}{h_0} \sin kl \\
								-e^{-\ii \phi } k \left(e^{\ii \theta } \frac{h_1}{h_0} \cos kl+k \sin kl\right) & k\left(e^{-\ii \theta}\frac{h_1}{h_0}\cos kl+k\sin kl\right)},\nonumber \\	
F_1(x)&=&\left(R_{L_1}(\lambda)g\right)(x), \ \ F_1'(x)=\frac{\dd}{\dd x} F_1(x).								
								}{Mmatice}
The solution exists if $\det M \neq 0$. The condition $\det M=0$ yields the equation (\ref{PTSepImpl}) for eigenvalues. 

Solutions $C_{\pm}$ have form $p_{\pm}(k,F_1,F_1'(0))/\det M,$ where expressions $p_{\pm} (k,F_1(0),F_1'(0))$ can obtained easily from (\ref{EqCpm}). If we consider $k$ for which $\det M \neq 0$ and $W(k)\neq 0$, then $C_{\pm}$ are bounded and estimates 
\eq{|C_{\pm}|\leq C \|g\|}
are valid for some constant $C$. $R_{L_1}(\lambda)$ is a compact operator and if we add rank one, i.e. also compact, operators $C_{\pm}(k)e_{\pm}$ we get $R_{L_{\phi}}(\lambda)$ which is then also compact for the fixed $k$. Whence, by the resolvent identity, $R_{L_{\phi}}(\lambda)$ is compact for all $\lambda \in \varrho(L_{\phi})$. 

This claim remains true also for $\phi=\pi/2$ and $\theta\neq 0,\pi$  because 
\eq{\det M = 2 k^2 \frac{h_1}{h_0} \sin\theta.  }

We can alternatively finish the proof by using \cite[III, Corollary 6.34]{Kato}. In order to prove that resolvent is compact it suffices to show that the resolvent set is non-empty, i.e. to find some $k$ for which $R_{L_{\phi}}(\lambda) \in \BH$.
\begin{flushright}
$\square$
\end{flushright}
\end{pf}

\section*{Acknowledgement}

I am thankful to D. Krej\v ci\v r\'ik and H. B\'ila for valuable discussions. This work is supported by M\v SMT 'Doppler Institute' project Nr. LC06002 and by the GA\v CR grant Nr. 202/07/1307.

\bibliographystyle{klunum}
\bibliography{references}

\end{article}
\end{document}